\documentclass[
reprint,
superscriptaddress,
amsmath,amssymb,
aps,
prx,
]{revtex4-2}

\usepackage{tabularx}
\usepackage{graphicx}
\usepackage{dcolumn}
\usepackage{upgreek}
\usepackage{mathrsfs}
\usepackage{setspace}
\usepackage{bm}
\usepackage[breaklinks=true,colorlinks=true,linkcolor=blue,urlcolor=blue,citecolor=blue]{hyperref}
\allowdisplaybreaks[4]

\usepackage{qcircuit}

\usepackage{bbm}
\usepackage{cancel}
\usepackage{comment}
\usepackage{physics}

\begin{document}

\title{Realization of an extremely anisotropic Heisenberg magnet in Rydberg atom arrays}


\author{Kangheun Kim}
\thanks{These authors contributed equally to this work}
\affiliation{Department of Physics, KAIST, Daejeon 34141, Republic of Korea}

\author{Fan Yang}
\thanks{These authors contributed equally to this work}
\affiliation{Center for Complex Quantum Systems, Department of Physics and Astronomy, Aarhus University, DK-8000 Aarhus C, Denmark}

\author{Klaus M{\o}lmer}
\affiliation{Niels Bohr Institute, University of Copenhagen, Blegdamsvej 17, DK-2100 Copenhagen, Denmark}

\author{Jaewook Ahn}
\email{jwahn@kaist.ac.kr}
\affiliation{Department of Physics, KAIST, Daejeon 34141, Republic of Korea}
	
	\begin{abstract} \noindent
		Strong mutual interactions correlate elementary excitations of quantum matter and plays a key role in a range of emergent phenomena \cite{winkler2006repulsively,fukuhara2013microscopic,preiss2015strongly,yan2019strongly,morvan2022formation}, from binding and condensation \cite{bottcher2020new} to quantum thermalization and many-body localization \cite{abanin2019colloquium}. Here, we employ a Rydberg quantum simulator to experimentally demonstrate strongly correlated spin transport in anisotropic Heisenberg magnets, where the magnon-magnon interaction can be tuned two orders of magnitude larger than the magnon hopping strength. In our approach, the motion of magnons is controlled by an induced spin-exchange interaction through Rydberg dressing \cite{yang2019quantum}, which enables coherent transport of a single Rydberg excitation across a chain of ground-state atoms. As the most prominent signature of a giant anisotropy, we show that nearby Rydberg excitations form distinct types of magnon bound states, where a tightly bound pair exhibits frozen dynamics in a fragmented Hilbert space, while a loosely bound pair propagates and establishes correlations beyond a single lattice site. Our scheme complements studies using resonant dipole-dipole interactions between Rydberg states, and opens the door to exploring quantum thermodynamics with ultrastrong interactions and kinetic constraints \cite{regnault2022quantum}.
	\end{abstract}
	
	\maketitle
	
\noindent
Quantum simulation of spin models has established a powerful tool for unraveling exotic many-body phases and dynamics \cite{bernien2017probing,zhang2017observation,gross2017quantum,kim2018detailed,ebadi2021quantum,scholl2021quantum,chen2023continuous}. As a pivotal process in quantum magnetism, the quasiparticle spin excitations (magnons) can propagate through the system by coherent spin exchanges that conserve the total magnetization \cite{auerbach1998interacting}. The inclusion of strong magnon-magnon interaction complicates the underlying spin transport, where the motion of different magnons cannot be separated \cite{bethe1931theorie,wortis1963bound,ganahl2012observation}. Similar correlated transport dynamics has been observed in various quantum systems, including ultracold atoms engineered by the superexchange mechanism \cite{fukuhara2013microscopic}, trapped atomic ions with phonon mediated spin-spin couplings \cite{kranzl2022observation}, and Rydberg atom arrays subjected to resonant dipole-dipole interactions \cite{scholl2022microwave}. These works aim to construct a spin-1/2 Heisenberg model, where the correlations can be tuned by the anisotropy of the XXZ-type Hamiltonian, defined as the strength of the magnon-magnon interaction relative to the spin-exchange rate. 

One of the biggest challenges in previous experiments was to acquire a very large anisotropy, for which the strongly correlated dynamics is constrained to flip-flops that conserve not only the total magnetization but also the number of domain-walls. This kinetic constraint is key to exotic non-ergodic dynamics, such as Hilbert space fragmentation \cite{regnault2022quantum} and quantum many-body scars \cite{turner2018weak}. In this work, we demonstrate an approach that can access such an extremely anisotropic regime on a neutral-atom quantum simulator, where ground-state atoms are off-resonantly dressed to a Rydberg state to induce an effective excitation exchange \cite{yang2019quantum}. As evidence of the large anisotropy, we show that the propagation of a single Rydberg excitation significantly slows down in the presence of a nearest-neighbor Rydberg excitation, due to the formation of a tightly bound state. While similar magnon bound states have been identified in systems with short-range interactions \cite{fukuhara2013microscopic} or moderate anisotropies \cite{kranzl2022observation}, the large long-range anisotropy in our work can further support a new type of bound states with a bond length beyond the nearest neighbor.

\begin{figure*}
	\centering
	\includegraphics[width=\linewidth]{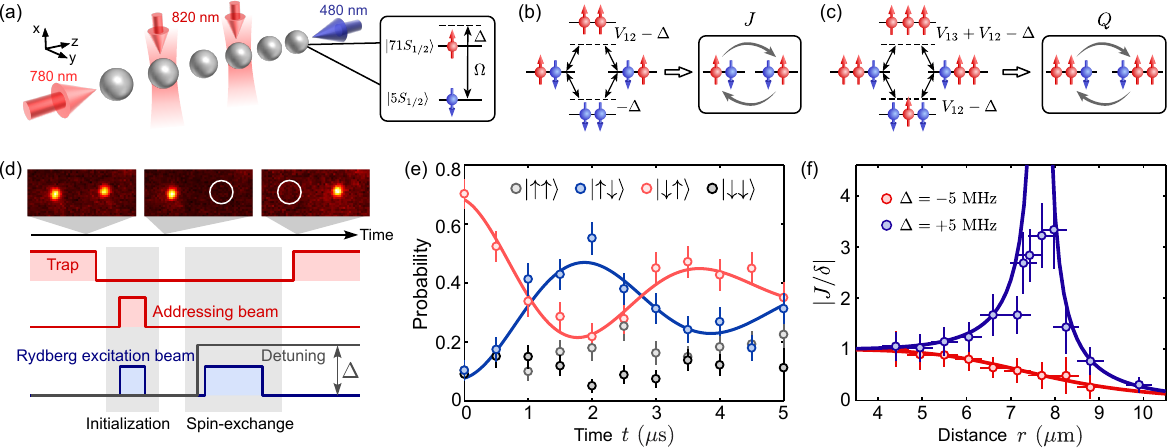}
	\caption{\textbf{Observation of spin-exchange dynamics in a Rydberg atom array.} (a) In the experiment, two counter-propagating lasers of $780~$nm and $480~$nm drive a two-photon transition, which couples the the ground state $|5S_{1/2},F=2,m_F=2\rangle$ to the Rydberg state $|71S_{1/2},m_J=1/2\rangle$ via an intermediate state $|5P_{3/2},F = 3, m_F= 3\rangle$. The trap laser of $820~$nm is reused as the individual addressing beam to provide site-dependent detunings through the a.c.-Stark shift. (b) and (c) illustrate perturbation mechanisms that induce two-body and three-body spin-exchange interactions, respectively. (d) Experimental sequence for probing the spin-exchange dynamics. For the state read-out, only atoms in the ground state $\ket{\downarrow}$ are detected, while Rydberg excitations $\ket{\uparrow}$ are detected as loss (indicated by white circles). (e) Measured spin-exchange dynamics between two atoms, where the population of the states $\ket{\uparrow\downarrow}$ and $\ket{\downarrow\uparrow}$ are fitted by damped sinusoidal functions. (f) Plot of $|J|/\Omega^2$ as a function of the distance $r$. We measure the oscillating population in $\ket{\downarrow\uparrow}$ at different interatomic distances for both positive and negative detunings, and fit the data to extract the interaction strength $J$.}
	\label{fig:fig1}
\end{figure*}
\vspace{10pt}
\noindent\textbf{Effective spin exchange in a Rydberg Ising model}\\
\noindent
Our experiments are carried out in a chain of $^{87}\rm Rb$ atoms initially trapped in an optical tweezer array [see Fig.~\ref{fig:fig1}(a)]. We use a two-photon excitation scheme to couple the ground state $\ket{\downarrow}=\ket{5S_{1/2},F=2,m_F = 2}$ to the Rydberg state $\ket{\uparrow}=\ket{71S_{1/2},m_J=1/2}$, which maps the system onto a spin-$1/2$ chain described by a tilted Ising Hamiltonian (taking $\hbar=1$, where $\hbar$ is the reduced Planck constant),
\begin{equation}
	\hat{H}_{\rm Ryd} = \frac{\Omega}{2} \sum_i  \hat{\sigma}_i^{x} - {\Delta}  \sum_i \hat{n}_i + \frac{1}{2}\sum_{i\neq j}V_{ij}\hat{n}_i \hat{n}_j.\label{eq:Ising}
\end{equation}
Here, $\hat{\sigma}_i^\alpha$ are Pauli matrices, $\hat{n}_i = |r_i\rangle\langle r_i|=(1+\hat{\sigma}_i^z)/2$ denotes the Rydberg-state projector, and $\Omega$ and $\Delta$ are the Rabi frequency and the detuning of the two-photon transition, respectively. The interaction strength $V_{ij}$ between Rydberg atoms at sites $i$ and $j$ takes the form $V_{ij}=C_6/r_{ij}^6$, where $r_{ij}$ is the distance between the atoms and $C_6>0$ is the van der Waals (vdW) coefficient. 

To understand the dynamics of this Rydberg Ising model, we decompose the original Hamiltonian into $\hat{H}_{\rm Ryd} = \hat{H}_0 + \hat{\Omega}_D$, where $\hat{H}_0$ is the diagonal part, and $\hat{\Omega}_D=(\Omega/2)\sum_i\hat{\sigma}_i^x$ is the off-diagonal driving term that can create or annihilate a single Rydberg excitation. If we label the eigenstates of $\hat{H}_0$ according to the total Rydberg excitation number $\hat{\mathcal{N}}_\mathrm{R}=\sum_i \hat{n}_i$, then $\hat{\Omega}_D$ only couples states where $\hat{\mathcal{N}}_\mathrm{R}$ changes by one. As a result, the coupling usually admixes different $\hat{\mathcal{N}}_\mathrm{R}$ subspaces. However, if the energy difference between adjacent blocks of $\hat{H}_0$ is much larger than the coupling strength $\Omega$, these subspaces become dynamically decoupled, and only states of the same $\hat{\mathcal{N}}_\mathrm{R}$ are coupled with each other via a perturbation process. This perturbation effect occurs predominantly at the second order and can be described by an effective Hamiltonian $\hat{H}_\mathrm{eff}$ (see Methods), which has a U(1) symmetry corresponding to the conserved Rydberg excitation number $\hat{\mathcal{N}}_\mathrm{R}$. Figure \ref{fig:fig1}(b) visualizes the perturbation process for two atoms, where states $\ket{\uparrow\downarrow}$ and $\ket{\downarrow\uparrow}$ are coupled by a spin-exchange interaction $J\left(\hat{\sigma}_1^+\hat{\sigma}_{2}^-+\hat{\sigma}_1^-\hat{\sigma}_{2}^+\right)$ between the ground state and the Rydberg state, with $\hat{\sigma}^\pm_n=\left(\hat{\sigma}^x_n \pm {\rm i}\hat{\sigma}^y_n\right)/2$. Crucially, the nonvanishing interaction strength $J = \Omega^2 V_{12}/4\Delta(\Delta-V_{12})$ is enabled by unequal energy differences between adjacent $\hat{\mathcal{N}}_\mathrm{R}$ sectors. These nonuniform level spacings arise from the vdW interaction and can lead to complicated density-dependent spin exchanges. For example, in a three-atom chain with the central site excited to the Rydberg state [see Fig.~\ref{fig:fig1}(c)], the spin exchange between the first and the third atom is described by a three-body interaction term $Q\left(\hat{\sigma}_1^+\hat{\sigma}_{3}^-\hat{n}_{2}+\hat{\sigma}_1^-\hat{\sigma}_{3}^+\hat{n}_{2}\right)$, where $Q = \Omega^2 V_{13} /4(\Delta-V_{12})(\Delta - V_{12} - V_{13})$ is the density-dependent coupling strength.

To observe these virtual spin-exchange processes, it is preferable to work in the weak dressing regime $\Omega\ll|\Delta|$, which, however, results in weaker interaction strengths. Concerning this trade-off, which could be relaxed by a larger Rabi frequency, our experiments are typically performed with $|\Delta/\Omega|\in [1.5,4]$. In this intermediate regime, we demonstrate that the U(1) symmetry is largely preserved and the deviation from the effective theory can be suppressed by a postselection measurement. Actually, we can accurately count Rydberg excitations in each experimental run by single-site resolved fluorescence imaging, which projects the spins to an exact microstate. Therefore, when exploring the dynamics of a specific $\hat{\mathcal{N}}_\mathrm{R}$ subspace, events subject to processes breaking the U(1) symmetry can be discarded, while only states remaining in the given symmetry sector are retained \cite{fukuhara2013microscopic}. This postselection scheme has a high success probability and shows good tolerance to imperfect state initialization.

\begin{figure*}
	\centerline{\includegraphics [width=\linewidth]{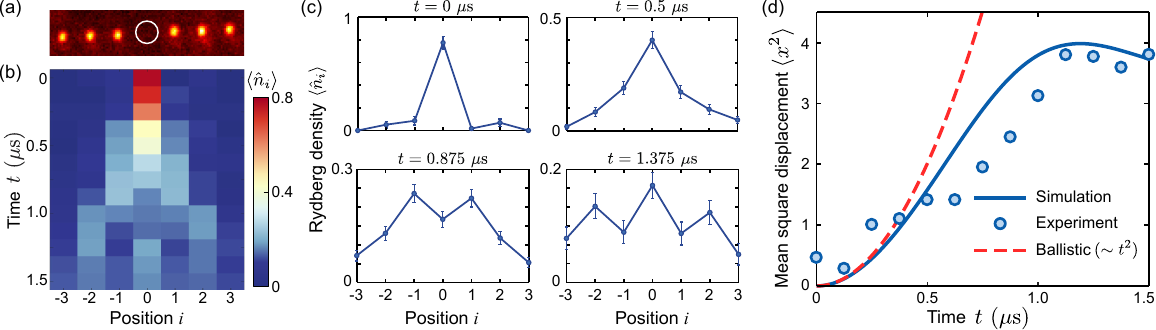}}
	\caption{\textbf{Quantum random walk of a single Rydberg excitation.} (a) Fluorescence image of the prepared initial state in the quench dynamics. To create the local Rydberg excitation (white circle), the atom at the central site is detuned by the addressing beam in a way that only this atom is excited by the global Rydberg beam (see Methods). (b) Evolution of the Rydberg density $\langle \hat{n}_i\rangle$. Here, we apply postselection and consider only data with a single Rydberg excitation. (c) Distribution of the Rydberg density $\langle \hat{n}_i\rangle$ at different times. (d) Mean square displacement $\langle x^2 \rangle$ as a function of time. The measured data, the calculated results, and the analytical ballistic estimations, are represented by the circles, the solid lines, and the dashed lines, respectively. The data shown here are state-preparation-and-measurement (SPAM) corrected with maximum likelihood estimation (see Methods).}
	\label{fig:fig2}
\end{figure*}
\vspace{10pt}
\noindent\textbf{Quantum walk of a single magnon}\label{sec:sec3}\\
\noindent
We first investigate the dynamics within the $\hat{\mathcal{N}}_R=1$ subspace of a single Rydberg excitation (magnon). The effective Hamiltonian for this symmetry sector is a simple XY model describing coherent hopping of a single magnon: $\hat{H}_{\rm eff} = \sum_{i< j} J_{ij} (\hat{\sigma}_i^+ \hat{\sigma}_j^- + \hat{\sigma}_i^- \hat{\sigma}_j^+) + \sum_i \mu_i \hat{n}_i$, where $J_{ij}= \Omega^2 V_{ij}/4\Delta(\Delta-V_{ij})$ is the rate of the effective spin exchange, and $\mu_i = -\Delta +2\delta +\sum_{j\neq i}J_{ij}$ is the on-site potential of the magnon with $\delta=\Omega^2/4\Delta$.

As a minimal yet nontrivial example, we begin with two sites and measure the spin-exchange process $\ket{\downarrow\uparrow}\leftrightarrow\ket{\uparrow\downarrow}$. To this end, two atoms are loaded into the tweezers and prepared in state $\ket{\downarrow\downarrow}$ via optical pumping. Then, the trap is turned off, and the first atom is addressed with a 820-nm laser, making it off-resonant with respect to the transition driven by the global Rydberg beam. The second atom is on-resonant and subsequently driven to the Rydberg state by a $\pi$-pulse, creating the desired initial state $\ket{\downarrow\uparrow}$. After that, the global Rydberg beam is significantly detuned to induce the effective spin exchange. The experimental sequence is shown in Fig.~\ref{fig:fig1}(d), and more details can be found in Refs.~\cite{kim2022quantum}. Figure \ref{fig:fig1}(e) depicts the characteristic oscillation dynamics measured with $\Omega = 2\pi\times 1.52 ~\rm MHz$, $\Delta = 2\pi\times 5 ~\rm MHz$, and $r=4.95~\rm \mu m$, where $r$ is the interatomic distance. It is clearly seen that the oscillation is approximately U(1) symmetric, as it mainly occurs in the single-excitation subspace, while states $\ket{\downarrow\downarrow}$ and $\ket{\uparrow\uparrow}$ are rarely populated. The oscillation frequency $\sim 0.80 ~\rm MHz$ drawn from the experiment agrees well with the perturbation analysis that gives $|J| \approx 0.78 ~\rm MHz$. Here, the damping of the coherent spin exchange is mainly caused by uncorrelated dephasings from the intermediate-state scattering, and the scheme is intrinsically robust against correlated dephasings from the laser phase noise.

We next measure the distance dependence of the interaction $J_{ij}=J(r_{ij})$ by varying the distance $r$ between the two atoms. As shown in Fig.~\ref{fig:fig1}(f), the measured potential perfectly matches the theoretical prediction $J_\pm(r) = {\delta}/{[(r/r_c)^6\mp1]}$, where $\pm$ denotes the sign of the detuning, and $r_c = (C_6/|\Delta|)^{1/6}$ is a characteristic length. For a negative detuning ($\Delta<0$), $J_-(r)$ is a soft-core potential that plateaus at $\delta$ for $r<r_c$ and decays with a vdW tail $\sim 1/r^{6}$, similar to the Rydberg-dressing induced interaction between ground-state atoms \cite{johnson2010interactions,henkel2010three,jau2016entangling,zeiher2016many,schine2022long,steinert2023spatially}. The potential for a positive detuning ($\Delta>0$) has a distinct behavior: while it has the same plateau value and asymptotic scaling, $J_+(r)$ diverges at $r=r_c$. This singularity is caused by the facilitation dynamics, where the condition $V_{i,i+1}=\Delta$ makes single-magnon states resonantly coupled with the two-magnon state $\ket{\uparrow\uparrow}$, leading to a breakdown of perturbation theory and the U(1) symmetry. In the facilitation regime, it has been shown previously that a small thermal fluctuation of atomic positions can lead to a strong Anderson localization, hindering the transport of the excitation \cite{marcuzzi2017facilitation}. In contrast, for the U(1) symmetric regime studied in this work, the plateau of the potential makes the dynamics insensitive to the fluctuation of interatomic distance, and a magnon is expected to be highly delocalized.

\begin{figure*}
	\centerline{\includegraphics [width=\linewidth]{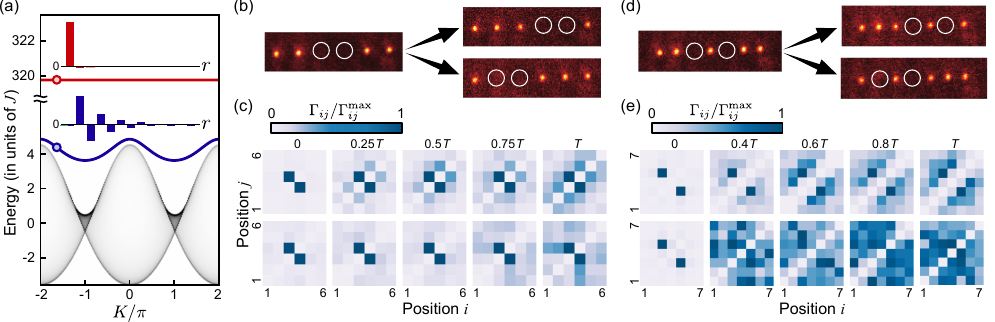}}
	\caption{\textbf{Correlated transport by magnon bound states.} (a) Two-excitation spectrum for $\Delta/\Omega=-3$ and $V_{i,i+1}/\Delta=-8$, where the red (blue) curve shows the dispersion relation of the tightly (loosely) bound pair outside the continuum of scattering states (shaded area, where the grayscale indicates the density of states). The wavefunction $\phi_K(r)$ is illustrated (bars) for bound states of the indicated momentum (circles). (b) Correlated transport of initial Rydberg excitations (3rd and 4th atom, marked with white circles), mediated by a next-nearest-neighbor hopping ($3\rightarrow5$; or $4\rightarrow2$). (c) Time evolution of the correlator $\Gamma_{ij}$ after preparation of nearest-neighbor pair of excitations ($T = 2.8\pi/J$). The upper panels are measured with $\Omega = 2\pi\times 2.54~\rm MHz$, $\Delta = 2\pi\times 12~\rm MHz$, and $r=7~\mu \rm m$, with a considerable hopping strength $Q\approx0.13~\rm MHz$. The lower panels are measured with a different detuning $\Delta = 2\pi\times -3.3~\rm MHz$, which leads to frozen dynamics due to a small $Q\approx0.01~\rm MHz$. (d) Correlated transport of initial Rydberg excitations (3rd and 5th site, marked with white circles), mediated by successive nearest-neighbor hoppings ($5\rightarrow6$, $3\rightarrow4$; or $3\rightarrow2$, $5\rightarrow4$). (e) Evolution of the correlator $\Gamma_{ij}$ after preparation of a next-nearest-neighbor pair of excitation ($T=1.7\pi/J$). The upper panels are measured with $\Omega = 2\pi\times2.06~\rm MHz$, $\Delta = 2\pi\times-3~\rm MHz$, and $r=4.95~\mu\rm m$, for which the initial state has a large overlap $\approx0.24$ with the bound state. The lower panels are measured with a larger lattice spacing $r=8.5~\mu\rm m$, where the initial state has a small overlap $\approx0.09$ with the bound state.}
	\label{fig:fig3}
\end{figure*}

To demonstrate that the magnon can exhibit robust quantum walk against atomic positional disorders, we now create a larger array containing 7 atoms with a spacing of $4.95~\rm nm$. In order to prepare the initial state $\ket{\downarrow\downarrow\downarrow\uparrow\downarrow\downarrow\downarrow}$, we apply the individual addressing beam to shift the detuning of the central site, followed by an adiabatic ramping of the global Rydberg beam, which only drives the atom at the center to the Rydberg state [Fig.~\ref{fig:fig2}(a)]. After the initialization, the addressing beam is turned off, and a red-detuned ($\Delta<0$) Rydberg driving field is applied to induce the effective dynamics. The propagation of the initial excitation can be traced by observing the evolution of the local Rydberg density $\langle\hat{n}_i\rangle$, as shown in Fig.~\ref{fig:fig2}(b), where an approximate light-cone wavefront can be identified. The staggered pattern of $\langle\hat{n}_i\rangle$ during the evolution is a clear evidence of the quantum interference [Fig.~\ref{fig:fig2}(c)], as opposed to the Gaussian distribution in a classical random walk. In the current system, the existence of uncorrelated dephasings will eventually destroy the coherence of the system and leads to a uniform steady distribution. To quantify the role of the dephasing, we extract the mean square displacement $\langle x^2 \rangle$ of the magnon [Fig.~\ref{fig:fig2}(d)], and find good agreement with the simulations based on the Haken-Reineker-Strobl (HRS) model \cite{haken1972coupled,haken1973exactly}, which includes both coherent magnon hoppings and on-site dephasings (with a rate $\gamma=2\pi\times 0.2~\rm MHz$). For a larger system, the HRS model predicts that the magnon will continue to spread with no steady-state distribution, but its motion has a quantum-classical crossover: while the initial propagation for $t<1/\gamma$ is governed by a ballistic transport ($\langle x^2 \rangle\propto t^2$), the spreading will gradually become diffusive with $\langle x^2 \rangle\propto t$. Such a scaling crossover can be identified in future experiments with increased system size.

\vspace{10pt}
\noindent\textbf{Dynamics of magnon bound states}\label{sec:sec4}\\
\noindent
Having explored the single-magnon dynamics, we proceed to the observation of correlated motions of multiple magnons. In the two-excitation subspace ($\hat{\mathcal{N}}_R=2$), neglecting the essentially uniform on-site potential, the effective Hamiltonian now reads
\begin{equation}
	\hat{H}_\mathrm{eff} = \sum_{i<j\neq k}Q_{ijk} \left(\hat{\sigma}_{i}^+ \hat{\sigma}_{j}^-\hat{n}_k + \hat{n}_k\hat{\sigma}_{i}^-\hat{\sigma}_{j}^+\right) +  \sum_{i<j} U_{ij}\hat{n}_i \hat{n}_{j},
	\label{eq:XXZ}
\end{equation}
where $Q_{ijk} = (G_{ijk}+G_{jik})/2$ is the density-dependent hopping strength with $G_{ijk} = \Omega^2 V_{ij} /4(\Delta-V_{ik})(\Delta - V_{ik} - V_{ij})$, and $U_{ij}=V_{ij}-4J_{ij}+ \sum_{l\neq i,j}(G_{lij}-J_{li})$ denotes the density interaction between magnons. Note that the density interaction $U_{ij}\sim V_{ij}$ is mainly from the zeroth-order Hamiltonian $\hat{H}_0$, while the exchange interaction $Q_{ijk}$ is induced by the second-order perturbation. This leads to an important characteristic that $|U_{ij}/Q_{ijk}|\sim (2\Delta/\Omega)^2\gg1$, which makes Eq.~(\ref{eq:XXZ}) a long-ranged, highly anisotropic Heisenberg model.

One direct consequence of this large anisotropy is the emergence of a family of magnon bound states. In an infinite spin chain, the two-magnon eigenstate $\ket{\psi_K}=\sum_{i\neq j}\psi_K(i,j)\hat{\sigma}_i^+\hat{\sigma}_j^+\ket{\downarrow\downarrow\cdots \downarrow}$ can be labeled by the center-of-mass momentum $K$, where the wavefunction can be factorized as $\psi_K(i,j) = e^{iKR} \phi_K(r)$ by introducing the center-of-mass position $R = (i+j)/2$ and the relative distance $r= i-j$ \cite{piil2007tunneling,valiente2009scattering,letscher2018mobile} . The bound state has a bounded wavefunction $\phi_K(\infty)\rightarrow0$, whose energy is isolated from the scattering continuum. Therefore, systems initially in the bound state remain localized in the relative coordinate, in stark contrast to the scattering state, where individual excitations propagate freely. Figure \ref{fig:fig3}(a) shows the energy spectrum and the bound-state wavefunction for a typical parameter $\Delta/\Omega=-3$ and $V_{i,i+1}/\Delta=-8$. The extremely large nearest-neighbor (NN) anisotropy $\xi_1=U_{i,i+1}/Q_{i-1,i,i+1}\approx 684$ in this case gives rise to a high-energy bound state (red curve), where magnons are tightly bounded at a relative distance $r=1$ (nearest neighbors) for all momenta. The strong density interaction also has a significant long-range effect absent in a short-range interacting system \cite{fukuhara2013microscopic}: the next-nearest-neighbor (NNN) anisotropy $\xi_2=U_{i,i+2}/Q_{i-1,i,i+2}\approx 4$ is also quite large, and can thus support a low-energy loosely bound state (blue curve), whose wavefunction $\phi_K(r)$ has a larger bond-length $r>1$. We will focus on these two types of bound pairs in the experiment, and expect that the same system gives rise to further varieties of bound states at larger anisotropy or in different lattice configurations.

To probe the correlated dynamics of the tightly bound Rydberg pair, we prepare an initial state $\ket{\downarrow\downarrow\uparrow\uparrow\downarrow\downarrow}$ in a 6-atom chain via an adiabatic anti-blockade excitation scheme, where the detuning for the center two atoms are swept across the resonant point $\Delta=V_{i,i+1}/2$. We then quench the system to a fixed detuning and measure the evolution of the two-site correlator $\Gamma_{ij}=\langle \hat{\sigma}_i^+ \hat{\sigma}_j^+ \hat{\sigma}_i^- \hat{\sigma}_j^-\rangle$. For a postive detuning $\Delta=2\pi\times 12~\rm MHz$, the observed correlation function propagates almost perfectly along the directions $j=i\pm1$ [see the upper panels of Fig.~\ref{fig:fig3}(c)], demonstrating that two Rydberg excitations move in a correlated manner as expected [see Fig.~\ref{fig:fig3}(b)]. In fact, the large NN anisotropy $\xi_1\approx -35$ in our experiment makes the total NN-Rydberg bonds $\hat{\mathcal{N}}_\mathrm{RR}=\sum_i \hat{n}_i\hat{n}_{i+1}$ another conserved charge. The tightly bound Rydberg pairs constitute the symmetry sector $(\hat{\mathcal{N}}_\mathrm{R}=2,~\hat{\mathcal{N}}_\mathrm{RR}=1)$, whose dynamics are governed by an NNN hopping term $Q\sum_i (\hat{\sigma}_{i}^+ \hat{\sigma}_{i+2}^-\hat{n}_{i+1} +\mathrm{H.c.})$. Here, the strength $Q=Q_{i,i+2,i+1}$ corresponds to the exchange process illustrated in Fig.~\ref{fig:fig1}(c), and determines the propagation speed of the tightly bound pair. To further confirm this analysis, we turn the detuning to a negative value $\Delta=2\pi\times-3.3~\rm MHz$, with which the single-magnon hopping strength $J=J_{i,i+1}$ remains unchanged, but the density-dependent hopping is significantly reduced ($Q=0.13~{\rm MHz} \rightarrow 0.01~\rm MHz$). Consistent with the theoretical prediction, the dynamics of the system becomes almost frozen within the time scale $T\sim 2\pi/J$ [see the lower panels of Fig.~\ref{fig:fig3}(c)], at which a single Rydberg excitation should already spread over the lattice. Note that the slight spreading of the correlator at late time is mainly caused by the imperfect state initialization rather than by excitation hopping. The frozen dynamics observed here is a clear signature of the Hilbert space fragmentation: while all tightly bound states $\ket{\cdots\uparrow_{i}\uparrow_{i+1}\cdots}$ share the local symmetry ($\hat{\mathcal{N}}_\mathrm{R}$ and $\hat{\mathcal{N}}_\mathrm{RR}$), they form dynamically disconnected Krylov subspaces of dimension 1 (frozen states). In fact, taking only NN vdW interactions into consideration (in accordance with a vanishing NNN hopping strength $Q$), the effective Hamiltonian can be mapped to a folded XXZ model \cite{de2019dynamics,yang2020hilbert,yoshinaga2022emergence,yang2023hibert}, where spin exchanges are constrained by the conservation of $\hat{\mathcal{N}}_\mathrm{RR}$, leading to a strongly fragmented Hilbert space in the thermodynamic limit.

Unlike the tightly bound state, which has a nearly flat band in most parameter regimes (corresponding to the frozen dynamics), the loosely bound pair displays a finite bandwidth and is therefore more mobile [Fig.~\ref{fig:fig3}(a)]. To observe the propagation of this longer-range bound state, we prepare a 7-site chain and excite the third and the fifth atom to the Rydberg level. We first choose a small lattice spacing of $4.95~\mu\rm m$ to achieve large anisotropies $\xi_1=539$ and $\xi_2\approx 1.24$, for which the produced initial state $\ket{\downarrow\downarrow\uparrow\downarrow\uparrow\downarrow\downarrow}$ has a considerable overlap ($\approx 0.24$) with the loosely bound state. The upper panels of Fig.~\ref{fig:fig3}(e) depicts the evolution of the experimentally extracted correlation function $\Gamma_{ij}$. In contrast to the tightly bound pair, whose transport is determined by an NNN hopping term, the correlated motion of the loosely bound pair is mediated by two successive NN hopping processes [Fig.~\ref{fig:fig3}(d)], as evident from the predominant spreading of $\Gamma_{ij}$ along the directions $i=j\pm2$. As a comparison, we then increase the interatomic distance to $8.5~\mu \rm m$, at which the NNN anisotropy $\xi_2\approx -0.52$ is too small to support the long-range bound state for most values of the momenta. In this regime, the observed correlator $\Gamma_{ij}$ rapidly spreads over the entire zone with no preferred propagation direction [see the lower panels of Fig.~\ref{fig:fig3}(e)], which suggests that the two Rydberg excitations are not bounded to each other but propagate freely \cite{kranzl2022observation}.
\begin{figure}
	\centering
	\includegraphics [width=\linewidth]{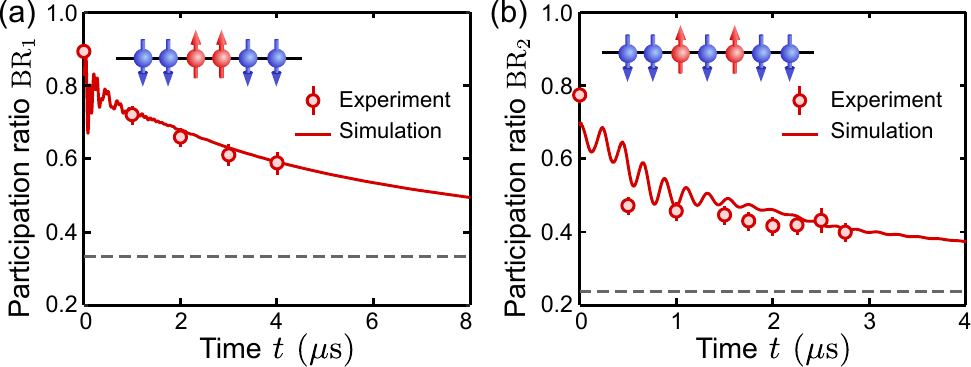}
	\caption{\textbf{Participation ratio for the tightly bound state (a) and the loosely bound state (b)}. The inset shows the initial spin configurations. The numerical simulation results are obtained by solving a quantum master equation, which includes the Rydberg decay, uncorrelated dephasings, as well as global laser phase noises. The simulation is averaged over 100 positional disorder realizations, and takes into account imperfect state initialization and detection.}
	\label{fig:fig4}
\end{figure}

To further confirm the existence of the bound state, we extract their participation ratios (BR) from the measured correlation map, where the ratios for the tightly bound state and the long-range bound state are defined as ${\rm BR}_1 = \sum_i\Gamma_{i,i+1}/\Gamma_\mathrm{tot}$ and ${\rm BR}_2 = \sum_i\Gamma_{i,i+2}/\Gamma_\mathrm{tot}$, respectively, with $\Gamma_\mathrm{tot} = \sum_{i<j}\Gamma_{ij}$. For the system size realized in our experiment, the reflection from the boundary can lead to a finite ${\rm BR}_1$ and ${\rm BR}_2$ even in the absence of magnon interactions. To estimate this finite-size effect and get a lower reference value for the participation ratio, we assume a uniform thermal distribution of the magnons with $\Gamma_{ij}=1/\Gamma_\mathrm{tot}$. As confirmed by Fig.~\ref{fig:fig4}, the measured ratio is much larger than this lower bound (dashed curves) during the free-magnon relaxation time $\sim 1/J$. Here, the damping of the bound pair at late time is mainly caused by the local dephasing. It is here worth pointing out that an atomic positional disorder may slow down the propagation of bounded magnons more easily than single magnons, because it contributes a large disordered binding interaction $U_{ij}$ (especially for the tightly bound pair). To account for the decoherence, the positional disorder, as well as other imperfections, we carry out full numerical simulations based on realistic experimental conditions and the original Rydberg Ising model (see Methods). This full simulation agrees very well with the experimental data (see Fig.~\ref{fig:fig4}) and suggests improving the coherence of the correlated spin-exchange dynamics in future studies.

\vspace{10pt}
\noindent\textbf{Conclusions and outlook}\\
\noindent
In conclusion, we have demonstrated a new approach to constructing the Heisenberg-type spin model in a Rydberg atom array. Different from previous schemes realized by dipolar exchange interaction and Floquet engineering \cite{scholl2022microwave}, our approach is based on Rydberg dressing of an Ising Hamiltonian, which can offer a large and widely tunable anisotropy. In the current experiment, we focused on the single-magnon and the two-magnon sector. By creating more excitations in a large-scale array, the system may allow exploration of emergent Hilbert space fragmentation \cite{yoshinaga2022emergence} and the Krylov-restricted thermalization of multiple magnons \cite{regnault2022quantum}. The scheme also allows dynamical engineering of spin transport, topological pumping protocols and programmable entanglement distributions \cite{yang2019quantum}. Generalizations to higher dimension could lead to richer physics. In particular, in a 2D lattice, the inclusion of a multicolor dressing field could enable application of a synthetic gauge flux \cite{wu2022manipulating}, which can give rise to topologically protected chiral motion of the magnon-bound state and holds promise for observation of a chiral spin liquid \cite{weber2022experimentally}.
	
\begin{acknowledgements} \noindent
This research was supported by Samsung Science and Technology Foundation (SSTF-BA1301-52) and National Research Foundation of Korea (2017R1E1A1A01074307). F. Yang and K. M{\o}lmer acknowledge the support from Carlsberg Foundation through the ``Semper Ardens'' Research Project QCooL and from the Danish National Research Foundation (DNRF) through the Center of Excellence ``CCQ'' (Grant No. DNRF156). We thank L. You, T. Pohl, A. E. B. Nielsen, H. Yarloo, H. Zhang, A. Cooper, and X. Wu for valuable discussions.
\end{acknowledgements}

\bibliography{reference}

\clearpage
\newpage

\setcounter{figure}{0}
\renewcommand{\figurename}{\textbf{Extended Data Fig.}}
\renewcommand{\tablename}{\textbf{Extended Data Table}}
\newcolumntype{b}{>{\centering\arraybackslash}X}
\newcolumntype{s}{>{\centering\arraybackslash\hsize=.5\hsize}X}
\renewcommand{\arraystretch}{1.5}

\section*{Methods}
	
\subsection*{Effective Hamiltonian of the system}
\noindent
The effective U(1) symmetric model can be constructed from the Schrieffer-Wolff (SW) transformation \cite{bravyi2011schrieffer}. Up to the second-order perturbation, the effective Hamiltonian is given by $\hat{H}_\mathrm{eff} = \hat{H}_0 + \hat{H}_\mathrm{eff}^{(2)}$ with
\begin{equation}
	\hat{H}_\mathrm{eff}^{(2)}=\hat{\mathcal{P}}\left(\frac{1}{2}[\hat{\mathcal{S}},\hat{\Omega}_D]\right)\hat{\mathcal{P}},\label{eq:eq3}
\end{equation}
where $\hat{\mathcal{S}}$ is a generator satisfying $[\hat{\mathcal{S}},\hat{H}_0]+\hat{\Omega}_D=0$, and $\hat{\mathcal{P}}$ projects out terms that do not conserve $\hat{\mathcal{N}}_\mathrm{R}$. Formally, the generator can be expressed as
\begin{equation}
	\hat{\mathcal{S}}=\mathrm{i}\frac{\Omega}{2}\sum_i\frac{\hat{\sigma}_i^y}{\Delta - \sum_{j\neq i}V_{ij}\hat{n}_j}.
\end{equation}
It is difficult to get an explicit effective Hamiltonian using the above expression. Therefore, we expand $\hat{\mathcal{S}}$ in orders of the Rydberg excitation number that can influence the spin flip of a single atom at the $i$-th site, i.e.,
\begin{eqnarray}
	\hat{\mathcal{S}} &=& (\mathrm{2i}/\Omega) \delta\sum_i \hat{\sigma}_i^y + (\mathrm{2i}/\Omega) \sum_{i\neq j}J_{ij}\hat{\sigma}_i^y\hat{n}_j \nonumber\\ && + (\mathrm{i}/\Omega)\sum_{i\neq j\neq k}(G_{ijk}-J_{ij})\hat{\sigma}_i^y\hat{n}_j\hat{n}_k +\cdots ,
\end{eqnarray}
where $\delta = {\Omega^2}/{4\Delta}$,
\begin{eqnarray}
	J_{ij} = \frac{\Omega^2V_{ij}}{4\Delta(\Delta-V_{ij})},\ G_{ijk} = \frac{\Omega^2V_{ij}}{4(\Delta-V_{ik})(\Delta-V_{ik}-V_{ij})}.\nonumber
\end{eqnarray}
The above expansion then leads to an effective Hamiltonian $\hat{H}_\mathrm{eff}^{(2)} = \hat{\mathcal{H}}_\textrm{1-body}+\hat{\mathcal{H}}_\textrm{2-body}+\hat{\mathcal{H}}_\textrm{3-body}+\cdots$, where
\begin{align}
	\hat{\mathcal{H}}_\textrm{1-body} &= \delta\sum_i\hat{\sigma}_i^z,\nonumber\\
	\hat{\mathcal{H}}_\textrm{2-body} &= \sum_{i\neq j}\frac{J_{ij}}{2} \left(\hat{\sigma}_{i}^+\hat{\sigma}_{j}^- + \hat{\sigma}_{i}^-\hat{\sigma}_{j}^+ -2\hat{\sigma}_i^z\hat{n}_j\right) \nonumber \\
	\hat{\mathcal{H}}_\textrm{3-body} &= \sum_{i\neq j\neq k}\frac{G_{ijk}-J_{ij}}{2} \left(\hat{\sigma}_{i}^+ \hat{\sigma}_{j}^- + \hat{\sigma}_{i}^-\hat{\sigma}_{j}^+ -\hat{\sigma}_i^z\hat{n}_j\right)\hat{n}_k,\nonumber
\end{align}
are the one-body self-energy shift, the two-body XXZ-type Hamiltonian, and the three-body XXZ term, respectively. The Hamiltonian can be further simplified by the substitution $\hat{\sigma}_i^z = 2\hat{n}_i -1$ in a given state sector. For the single-magnon sector ($\hat{\mathcal{N}}_R=1$), the quadratic term $\hat{n}_i\hat{n}_j$ can be neglected, which leads to the XY model given in the main text. For the two-magnon sector ($\hat{\mathcal{N}}_R=2$), the cubic term $\hat{n}_i\hat{n}_j\hat{n}_k$ can be discarded, and the resulting Hamiltonian can be mapped to Eq.~(\ref{eq:XXZ}). For a general multi-magnon case, the dynamics is governed by a folded XXZ model exhibiting the HSF \cite{yang2023hibert}.


\subsection*{Experimental setup and procedure} 

\begin{figure*}
	\centerline{\includegraphics [width=\linewidth]{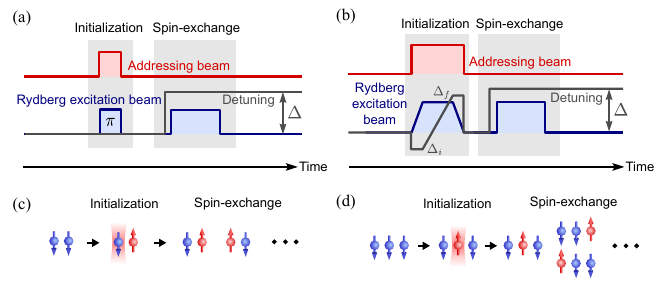}}
	\caption{\textbf{Experimental Sequence.} (a) Sequence for the two-atom experiment illustrated in (c). (b) Sequence for the quantum walk [illustrated in (d)] and the bound-state experiments.}\label{fig:picture}
	\label{figapp1}
\end{figure*}

\noindent
The experimental setup of our system is a Rydberg quantum simulator using a neutral atom array of $^{87} \rm Rb$ atoms, similar to our previous experiments \cite{kim2022quantum}. The atomic ensembles are cooled and gathered inside a magneto-optical trap (MOT), while the single atoms are trapped inside a 820-nm optical tweezer array of $1~\rm mK$ depth and sub-Doppler cooled to $\sim 35~\mu \rm K$ with polarization gradient cooling. Atoms are then optically pumped to $\ket{\downarrow} = \ket{5S_{1/2},F=2,m_F=2}$. After the ground state preparation, traps are turned off and the atoms are operated to the Rydberg state $\ket{\uparrow} = \ket{71S_{1/2},m_J=1/2}$ with the two Rydberg beams of 780-nm (homemade ECDL) and 480-nm (TA-SHG Pro of Toptica) with two photon transition of intermediate detuning of $\Delta_I = 2\pi \times 660~\rm MHz$ from the intermediate state $\ket{m} =\ket{5P_{3/2},F=3,m_F=3}$. Quantum operation is performed by a series of Rydberg and addressing laser pulses. After the quantum operation, atoms are trapped again by turning on the optical tweezer, and atoms in the Rydberg states are anti-trapped from the tweezer. The remaining atoms are imaged with the electron-multiplied charged coupled device (EMCCD, iXon Ultra 888 of Andor) by illuminating the imaging beam. By distinguishing the fluorescence of background and trapped atom, we could determine the internal state of each individual atom.

The optical tweezer trap and the addressing beam for the state initialization use the same 820-nm laser drived from Ti:Sapphire oscillator (TiC of Avesta) pumped by a 532-nm laser (Verdi G18 of Coherent). The laser beam passes an acousto-optic modulator (AOM) and is split into zeroth and first order beams. The first order beam is sent to the spatial light modulator (SLM, ODPDM512 of Meadowlark optics), and the optical tweezer array of target and reservoir traps is formed and rearranged with real-time calculation Gerchberg-Saxton weighted (GSW) algorithm with GPU (Titan-X Pascal of NVIDIA). The phase for atom arrays are calculated with a 4 times larger array zero-padded to the initial phase to achieve resolution less than the trap size~\cite{kim2019gerchbergsaxton}. The zeroth order beam propagates along a different path passing an additional AOM and followed by an acousto-optic deflector (AOD, DTSXY-400-820 of AA Opto-Electronic) which is used to address the target atom. This 820-nm addressing beam is off-resonant to the $5S\rightarrow 5P$ transition, inducing an a.c.-Stark shift to the target-atom Rydberg transition.

The quantum operation is programmed using a delay generator (DG645 of Stanford Research Systems) and an arbitrary waveform generator (AWG, XRF Agile RF Synthesizer of Moglabs), controlling AOMs of both the addressing beams and the Rydberg beams. The sequence is depicted in Fig.~\ref{fig:fig1}(d) of the main text, and a more detailed one is given in Extended Data Fig.~\ref{figapp1}. The sequence is divided into two parts: an initialization process driving the target atoms to Rydberg states, and the spin-exchange process inducing the many-body quench dynamics. For the two-atom experiment, the initial state is prepared by addressing one of the atoms to make it off-resonant to the Rydberg beams and applying a resonant $\pi$ pulse to the other atom [see Extended Data Fig.~\ref{figapp1}(a) and Fig.~\ref{figapp1}(c)]. For all other experiments, the target atoms are addressed, and the Rabi frequency $\Omega$ and the detuning $\Delta$ of the global Rydberg beams are adiabatically swept according to the following sequence: (1) $0~\mu s\rightarrow 0.1~\mu s$, $(0,\Delta_i)\rightarrow (\Omega_{\rm exp},\Delta_i)$ (2) $0.1~\mu s\rightarrow 0.9~\mu s$, $(\Omega_{\rm exp},\Delta_i)\rightarrow (\Omega_{\rm exp},\Delta_f)$, and (3) $0.9~\mu s\rightarrow 1~\mu s$, $(\Omega_{\rm exp},\Delta_f)\rightarrow (0,\Delta_f)$ as depicted in Extended Data Fig.~\ref{figapp1}(b), where $\Omega_{\rm exp}$ is the Rabi frequency used in the spin-exchange step. The values of these parameters are summarized in Extended Data Table~\ref{table2}. With the above initialization, the addressed target atom is adiabatically excited to the Rydberg state [see Extended Data Fig.~\ref{figapp1}(d)].

\subsection*{Experimental parameters and measured values} 
\noindent
The experimental parameters are given in the following tables. Extended Data Table~\ref{table1} shows the parameters and measured values for the two-atom spin-exchange dynamics, where $\Delta$ is the detuning for the spin exchange, $r$ is the distance between the two atoms, $\Omega$ is the Rabi frequency, and $J$ is the spin-exchange frequency fitted from each experiment, e.g., from the data in Fig.~\ref{fig:fig1}(e) of the main text. The vdW interaction strength $V = C_6/r^6$ is determined by the distance $r$ with $C_6 = 2\pi \times 1023~{\rm GHz}\cdot \mu {\rm m}^{-6}$ corresponding to the Rydberg state $\ket{71S_{1/2}, m_J = 1/2}$ used in the experiment \cite{weber2017calculation}. The values of $\Omega$ and $J$ are fitted to the expression $P=a+b\cos(2\pi\times c\times t)\times \exp(-t/d)$ with unknowns $a$, $b$, $c$, $d$ and probability $P$ of the initial state, where $\Omega/2\pi$ and $J/4\pi$ corresponds to $c$. The errors in $r$, which is plotted in Fig.~\ref{fig:fig1}(f) of the main text, has the same value $0.3~\mu \rm m$ for all distances, which is limited by the resolution of the image plane, where the beam waist is about $\sim 1.2~\mu \rm m$ and the resolution is $\sim 0.3~\mu \rm{ m }=1.2/4 ~\mu \rm m$ because of the zero-padding. Extended Data Table~\ref{table2} shows the experimental parameters for the rest of the experiments. Here, $\Omega_{\rm exp}$ is the Rabi frequency for both spin-exchange dynamics experiment and the maximum Rabi frequency for the quantum annealing in the initial state preparation, $\Delta_A$ is the detuning applied on the target atom by the addressing beam (two values respectively for the left and the right atom in the two-magnon experiments), $\Delta_i$ and $\Delta_f$ is the initial and final detuning respectively for the detuning sweep of the state initialization, and $\Delta_{\rm exp}$ is the detuning for the spin-exchange quench dynamics.

\begin{table*}
	\caption{Experimental parameters and measured values for the two-atom experiment}
	\centering
	\begin{tabularx}{\textwidth}{sb|bb|b}
		\hline \hline
		$\Delta$~(MHz) & $r$~($\mu$m) & $\Omega/2\pi$~(MHz) & $J/2\pi$~(MHz) & $2\pi J/\Omega^2$~$\left({\rm MHz}^{-1}\right)$\\
		\hline
		& 4.4  &  1.52(5) & 0.132(9) & 0.06(1) \\
		& 4.95 &  1.52(5)  & 0.128(8) & 0.06(1) \\
		& 5.5  &  1.52(5)  & 0.143(6) & 0.06(1) \\
		& 6.05 &  1.52(5)  & 0.154(6)&0.07(2)\\
		& 6.6 & 1.52(5)  & 0.21(2) &0.09(2) \\
		& 7.15 & 1.52(5)  & 0.21(1) & 0.09(2)\\
		+5& 7.29 &  1.52(5)  & 0.33(3) &0.14(3) \\
		& 7.43 &  1.52(5)  & 0.35(2) & 0.15(3) \\
		& 7.7 & 1.52(5)  & 0.40(3) & 0.17(3) \\
		& 7.98 & 1.52(5)  & 0.42(5) & 0.18(4) \\
		& 8.25 & 1.52(5)  & 0.18(3) & 0.08(3)\\
		& 8.8 & 1.52(5)  & 0.10(1) & 0.04(1) \\
		& 9.9 & 1.52(5)  & 0.039(6) & 0.017(8)\\
		\hline
		& 4.4 &  1.86(7) &0.161(2)&0.05(2) \\
		& 4.95 & 1.86(7) & 0.143(1) &0.04(2)\\
		& 5.5 & 1.52(5)  &0.095(8) & 0.04(1) \\
		& 6.05 & 1.52(5)  &0.086(9) & 0.04(1) \\
		-5 & 6.6 & 1.52(5)  & 0.068(6) & 0.03(1) \\
		& 7.15 & 1.52(5)  &0.06(1)& 0.03(1) \\
		& 7.7 &1.86(7)& 0.08(2) & 0.02(1) \\
		& 8.25 & 1.91(9) & 0.08(2) & 0.02(2) \\
		& 8.8 &  1.91(9)  & 0.04(1) & 0.01(1) \\
		\hline \hline
	\end{tabularx}
	\label{table1}
\end{table*}

\begin{table*}
	\caption{Experimental parameters for the quantum-walk and the bound-state experiments}
	\centering
	\begin{tabularx}{\textwidth}{b s s s s s s}
		\hline \hline
		Experiment & $r$~($\mu$m) & $\frac{\Omega_{\rm exp}}{2\pi}$~(MHz) & $\frac{\Delta_A}{2\pi}$~(MHz) &  $\frac{\Delta_i}{2\pi}$~(MHz) & $\frac{\Delta_f}{2\pi}$~(MHz) & $\frac{\Delta_{\rm exp}}{2\pi}$~(MHz)\\
		\hline
		Single-magnon quantum walk& 4.95 & 2.54& -15.8 & +5 & +30 & -5 \\
		\hline
		Tightly bound state correlated transport& 7 &  2.54 & -20.3/-18.6 & +10 & +35 & +12\\
		Tightly bound state frozen motion& 7 &2.54 &-20.3/-18.6 &+10 & +35 & -3.3 \\
		\hline
		Loosely bound state correlated transport & 4.95  & 2.06& -7.4/-5.4 & +3 & +15 & -3 \\
		Loosely bound state free propagation & 8.5 &2.06 & -7.4/-5.4 &  +3& +15 & -3 \\
		\hline \hline
	\end{tabularx}
	\label{table2}
\end{table*}

\begin{table*}
	\caption{Experimental errors and its treatment to numerical simulations}
	\centering
	\begin{tabularx}{\textwidth}{s b}
		\hline \hline
		Error source & Treatment \\
		\hline
		Individual dephasing& $\mathcal{L}_{\rm ind} = \sum_{j=1}^{N}{\left (L_j \rho L_j^{\dagger} - \frac{1}{2}\left \{L_j^{\dagger}L_j, \rho  \right \}  \right )}$ with $L_j=\sqrt{\gamma_{\rm ind}/2}\hat{n}_j$ and $\gamma_{\rm ind} \approx 2\pi \times 0.2\rm ~ MHz$ \\
		Collective dephasing& $\mathcal{L}_{\rm col} = L_0 \rho L_0^{\dagger} - \frac{1}{2}\left \{L_0^{\dagger}L_0, \rho  \right \}  $ with $L_0=\sqrt{\gamma_{\rm col}/2}\sum_{j=1}^N\hat{n}_j$ and $\gamma_{\rm col} \approx 2\pi \times 0.4\rm ~ MHz$ \\
		Finite temperature of atoms& Monte Carlo simulation with positional fluctuation where $\sigma_\mathrm{r} \approx 0.1~\mu\rm m$ (radial) and $\sigma_\mathrm{a} \approx 0.3~\mu\rm m$ (axial)\\
		$P(g|r)$ measurement error & $P(g|r) = 1-\exp(-t_{\rm trap}/t_1)$ with Rydberg decay time $t_1 = 43(15) ~\mu \rm s$ \\
		$P(r|g)$ measurement error & $P(r|g)=P_{\rm recap}(t_{\rm trap})$ where $P_{\rm recap}$ is the release and recapture probability curve \\
		\hline \hline
	\end{tabularx}
	\label{table3}
\end{table*}

\subsection*{Experimental imperfections and numerical simulations}
\noindent
Full numerical simulations in Fig.~\ref{fig:fig4} of the main text take the experimental errors into consideration. Extended Data Table~\ref{table3} shows types of experimental imperfections and its treatment in the numerical simulations. The dominant error in the dressing scheme is the uncorrelated individual dephasing mainly due to the spontaneous decay from the intermediate state, vdW interaction fluctuation due to the finite temperature of the atom, as well as the state-measurement error. The collective dephasing mainly induced by the laser phase noise does not have a significant role on the dynamics because of the decoherence-free feature of the effective model \cite{wu2022manipulating}. Both individual and collective dephasings are treated with the Lindblad master equation ${d\rho}/{dt} = -{\mathrm{i}}\left [H, \rho  \right ] + \mathcal{L}_{\rm ind}(\rho) + \mathcal{L}_{\rm col}(\rho)$ \cite{lee2019coherent}, where the superoperator $\mathcal{L}_{\rm ind}$, $\mathcal{L}_{\rm col}$ denotes the individual (on-site) and the collective phase noise, respectively. The individual dephasing rate $\gamma_{\rm ind} \approx 2\pi\times 0.2~\rm MHz$ was fitted from the three level model of $\ket{g}$, $\ket{r}$ and the intermediate state $\ket{m}$. The collective phase noise was fitted from the single-atom Rabi oscillation by fixing $\gamma_{\rm ind}$, and its value is $\gamma_{\rm col}\approx 2\pi\times 0.4~\rm MHz$. The temperature of the atomic thermal motion $T_{\rm atom} = 34.27(5) ~\mu\rm K$ was measured using release and recapture method. With the temperature, we could calculate the motional variation of atom with a standard deviation $\sigma_i = \sqrt{k_BT/(m\omega_i^2)}$ of the position  for the trap frequency $\omega_i$. In the simulation, the average effect of such an atomic positional disorder was evaluated with the Monte-Carlo method. The radial and longitudinal position standard variations are $\sigma_\mathrm{r} \approx 0.1~\mu \rm m$ and $\sigma_\mathrm{a} \approx 0.3~\mu \rm m$ respectively. The detection error was considered similar to \cite{de2018analysis}, where the dominant portion of the conditional error probability $P(g|r)$ is due to the Rydberg decay and the dominant portion of $P(r|g)$ is due to a finite temperature of the atom. The former is calculated with $P(g|r) = 1-\exp(-t_{\rm trap}/t_1)$, where $t_{\rm trap}$ is the time when the trap is turned off, and the Rydberg lifetime $t_1=43(15)~\mu\rm s$ is measured with an additional Ramsey experiment \cite{levine2018highfidelity}. The latter probability $P(r|g)=P_{\rm recap}(t_{\rm trap})$ is obtained from the release and recapture probability curve.

\end{document}